\documentclass[twocolumn,prb,showpacs,multicol,amsmath,amssymb]{revtex4}
\usepackage[dvips]{graphicx}
\usepackage{graphicx}
\usepackage{dcolumn}
\usepackage{bm}
\usepackage{graphics}
\usepackage{epsfig,color}
\usepackage[normalem]{ulem} 
\usepackage{soul}

\newcommand{\be}{\begin{equation}}
\newcommand{\ee}{\end{equation}}
\newcommand{\bea}{\begin{eqnarray}}
\newcommand{\eea}{\end{eqnarray}}

\newcommand{\bS}{\bf S}

\begin{document}
\title{Focusing on the dynamics of the entanglement in spin junction}
\author{Z. Saghafi, Z. Shadman, E. Hosseini Lapasar, S. Mahdavifar}
\email[]{smahdavifar@gmail.com}
\affiliation{ Department of Physics, University of Guilan, 41335-1914, Rasht, Iran}
\date{\today}

\begin{abstract}
We study the dynamics of entanglement in the one-dimensional spin-1/2 $XY$  model in the presence of a transverse magnetic field. A pair of spins are considered as an open quantum system, while the rest of the chain plays the role of the environment. Our study focuses on the pair of spins in the system, the edge spins, and the environment. 
It is observed that the entanglement between the pair of spins in the system decreases and it can transfer to the rest of the spins. For a value of anisotropy leading to the Ising model, the entanglement is completely back to the system by passing time. On the other hand, the entanglement can only be seen under certain conditions between edge spins  of the system and the environment. The pair of spins on the edge will be entangled very quickly and it will  disappear after a very short time. A pair of spins far from the system was chosen to examine the behavior of entanglement in the environment. As expected, the transmission of entanglement from the system to the environment takes notable time.

\end{abstract}
\maketitle

\section{Introduction}\label{sec1}

The coupling of the system to a noisy environment \cite{1} often disturbs the experimental investigation and the control of characteristic quantum properties of physical systems \cite{2}. Some quantum resources \cite{3}' \cite{4}, such as the quantum entanglement \cite{5}, the quantum correlation \cite{6}' \cite{7}, and the quantum coherence \cite{8}' \cite{9} happen because of noisy environment. They are entirely delicate in the presence of environment so that can be destroyed or generated \cite{10}. 

The approach of studying the dynamics of an open system can be thoroughly different depending on the specific system being examined; in the case of intricate discussion on important properties such as Markovian dynamics have involved many experimental scientists \cite{11}. 
Along the same line, some quantum properties of open systems are irreversible losses. Surprisingly, the most fascinating characteristic of entanglement is to revive by time evolution \cite{12,13,14}. 

Many open systems exhibit memory effect to revive entanglement. Creating an excitation in a localized entangled state, the evolution of entanglement generated by dynamics of system is investigated \cite{16}. The flow of entanglement between two cavities interacting with reservoirs is explored and found that as the entanglement of the system is suddenly decreased, it is suddenly and necessarily increased in reservoirs  \cite{17}. It has been shown that through decoherence process induced by a local interaction Hamiltonian, initial correlations in a composite environment can lead to a nonlocal open system dynamics exhibiting strong memory effects although the local dynamics is Markovian \cite{18}. 

While most works focus on dynamics of entanglement in systems \cite{15,19,21,21-1}, it is also interesting to consider its behavior in the junction point of system-environment. 
 Firstly studied by Ratner and Aviram in 1974  molecular junction has attracted a lot of attention in the last two decades\cite{22}. Transporting electron through junction which is used to construct simple electron device has unceasingly developed in both theoretical and experimental perspective \cite{23}. However, most studies have explored transporting electron through metal-molecule-metal junction \cite{24,25,26}, there is an investigation on one dimensional nano-material with two metal-molecule interfaces called single molecular junction \cite{27}. 

\begin{figure}[t]
\centerline{\psfig{file=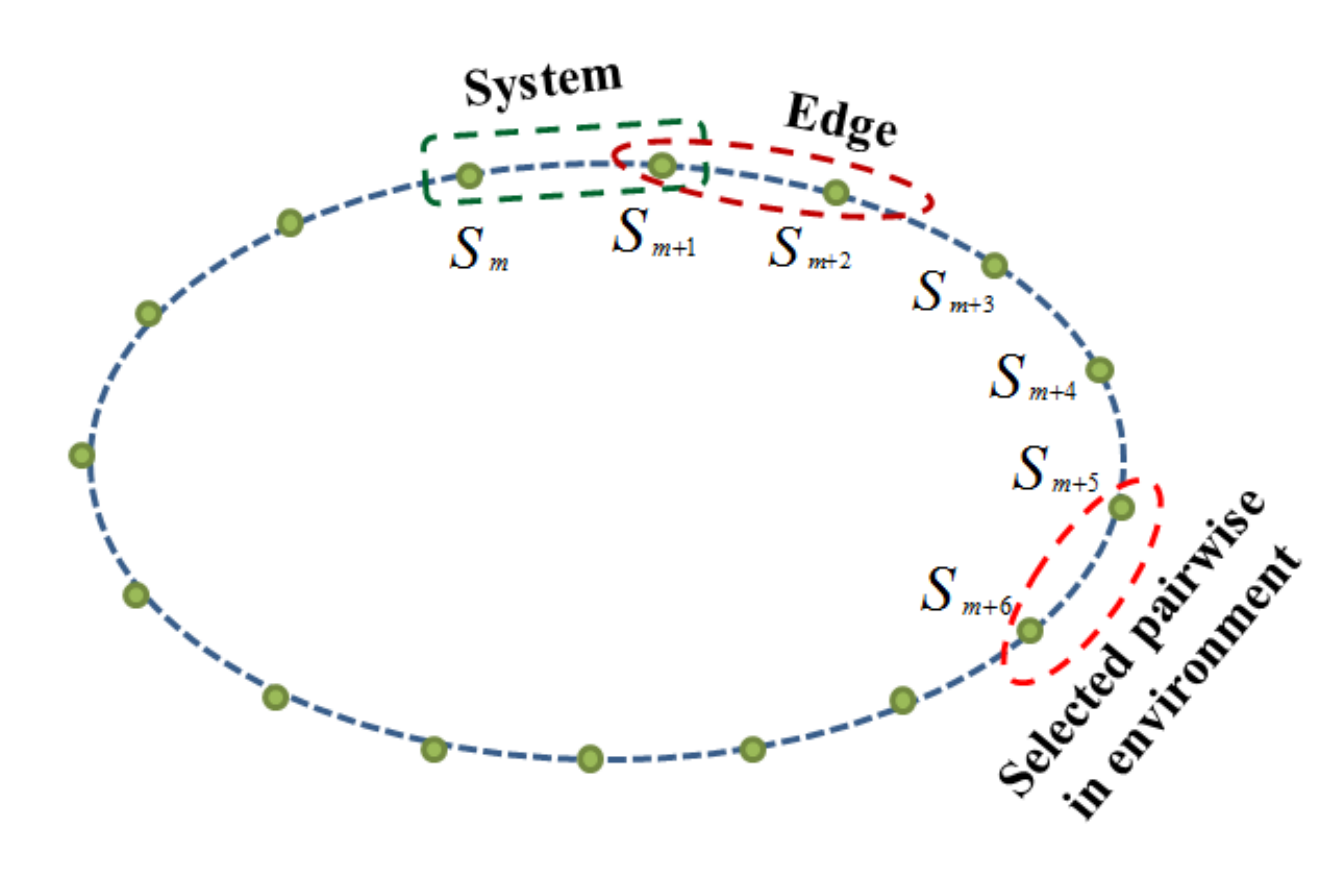,width=3.3 in}}
\caption{The symbolic diagram of a two-spin system and its environment.}\label{fig1}
\end{figure}

Recently, the dynamics of an open quantum system containing a pair of nearest neighbor spins coupled to a spin-1/2 XX chain environment  is studied\cite{15}.  It is found that the dynamical phase transition from the Markovian to the non-Markovian regime occurs by increasing the three-spin interaction. Here, we consider the spin-1/2 anisotropic XY model in the presence of a transverse magnetic field. Selecting the nearest neighbor pair spins as an open quantum system, the rest of the anisotropic chain plays the role of the environment. We consider a special initial state where the nearest  neighbor pair spins of the system are completely entangled, while no quantum correlations exist between other pair spins. First, using the analytical fermionization technique, the Hamiltonian is diagonalized. Then the time dependent quantum state is  obtained explicitly. Using the time dependent quantum state, the entanglement between every arbitrary nearest neighbor pair spins of the chain model is calculated as a function of the anisotropy, transverse field and the time.  We study the propagation of the entanglement from the system into the environment. Specially, we focus on the nearest neighbor pair spins on edges of the system and the environment (please see Fig.~\ref{fig1}). Depends on the anisotropy parameter, we report various behaviors of the entanglement between arbitrary nearest neighbor pair spins.

The paper is organized as follows. In the next section, we introduce the anisotropic spin-1/2 XY model and find an analytical form for the
entanglement and the time dependent quantum state of the system. In Sec. III, analytical results are presented. Finally, we summarize our results in Sec.
IV.       

\section{THE MODEL}\label{sec2}

The Hamiltonian of an anisotropic spin-1/2 chain in the presence of a magnetic field is written as\cite{16}
\begin{eqnarray}
{\cal H}&=&-J \sum_{n=1}^{N}[(1+\gamma){S}_{n}^x {S}_{n+1}^x+(1-\gamma){S}_{n}^y {S}_{n+1}^y]  \nonumber\\
  &-&h\sum_{n=1}^{N}{S}_{n}^{z}~,  
\label{Hamiltonian s}
\end{eqnarray}
where $S_{n}$ is the spin-1/2 operator on the $n$-th site, $J$ denotes the exchange coupling constant, $h$ is the applied magnetic field, and  $\gamma$ is the anisotropic parameter. Special values of $\gamma$ show various models. $\gamma=0$ and $\gamma=1$ represent isotropic XX and Ising model respectively. In the region $0<\gamma<1$, the exchange interaction along the x and y-direction will be ferromagnetic while in the range $\gamma>1$, we have ferromagnetic exchange interaction along the x-direction and antiferromagnetic exchange interaction along the y-direction. \\
The chain is considered in the thermodynamic limit with periodic boundary condition. One should note that the model has $U(1)$ symmetry at  $\gamma=0$. In contrast, applying $\gamma$ the symmetry breaking is observed. We briefly indicate steps of diagonalization of the model.  the one-dimensional spin-1/2 $XY$ chain reduces to  the one dimensional spinless fermions under Jordan-Wigner transformations \cite{35}
\begin{eqnarray}
{S}_{n}^{+}&=&a_{n}^{\dagger} \exp(i\pi\sum_{l<n}a_{l}^{\dagger}a_{l})~,  \nonumber\\
{S}_{n}^{-}&=&a_{n} \exp(-i\pi\sum_{l<n}a_{l}^{\dagger}a_{l})~, \nonumber\\
{S}_{n}^{z}&=&a_{n}^{\dagger}a_{n}-\frac{1}{2}~.
\end{eqnarray}
The Hamiltonian obtained as
\begin{eqnarray}
{\cal H}_{f}&=& \frac{-J}{2} \sum_{n}(a^{\dag}_{n}a_{n+1}+a^{\dag}_{n+1}a_{n} \nonumber \\
&+&\gamma (a^{\dag}_{n}a^{\dag}_{n+1}-a_{n}a_{n+1}))-h\sum_{n} a^{\dag}_{n}a_{n}~.
\end{eqnarray}
Applying a Fourier transformation as
\begin{eqnarray}
a_{n}^{\dag} = \frac{1}{\sqrt{N}}\sum _{j=1}^{N} e^{ikn} a_{k}^{\dag}, 
\end{eqnarray}
and Bogoliubov transformation \cite{36}
\begin{eqnarray}
\beta_{k}^{\dag}=cos(k) a_k^{\dag}-i sin(k) a_{-k},
\end{eqnarray}
the Hamiltonian will be diagonalized as
\begin{eqnarray}
{\cal H}_{f}=\sum_{k}\varepsilon(k) (\beta_{k}^{\dagger} \beta_{k}-\frac{1}{2}),
\label{Hamiltonian d}
\end{eqnarray}
where the energy spectrum is
\begin{eqnarray}
\varepsilon(k) &=& \sqrt{a(k)^2+4b(k)^2}, \nonumber\\
a(k)&=&-J \cos(k)-h, \nonumber\\
b(k)&=&\frac{J \gamma}{2} \sin(k).
\end{eqnarray}
In the following, we pay particular attention to the spreading of the pairwise entanglement by the effect of $\gamma$ and $h$. To reach this purpose, the initial state is considered such that two spins on sites $m$ and $m'$ are maximally entangled and the rest are disentangled at $t=0$.
\begin{eqnarray}
|\psi(t=0) \rangle&=&\frac{1}{\sqrt{2}}(a_m^{\dag}|0\rangle+e^{i \phi}a_{m'}^{\dag}|0\rangle) \nonumber\\
&=&\frac{1}{\sqrt{2N}}\sum_{k}(e^{ikm}+ e^{i(km'+\phi)})a_k^{\dag}|0\rangle~,
\end{eqnarray}
where $|0\rangle$ is the vacuum state of the original fermions, i.e. $a_k |0\rangle=0$ and $\phi$ is the phase factor. It is useful to note that although $a_k |0\rangle=0$, but $\beta_k |0\rangle\neq0$. The relation between the vacuum state of the original fermions and the vacuum state of the Bogoliubov operators can be shown as\cite{37}
\begin{eqnarray}
|0\rangle=\prod_k{(\cos(k)+i \sin(k) \beta^{\dag}_{k}\beta^{\dag}_{-k})}|\Omega\rangle,
\end{eqnarray}
where $|\Omega\rangle$ is the vacuum of the Bogoliubov operators. The challenging calculation happens here which explained with details in the index. Then, applying the time evolution operator as
\begin{eqnarray}
U(t)=e^{-it\sum_{k}\varepsilon(k) (\beta_{k}^{\dagger} \beta_{k}-\frac{1}{2})/\hbar},
\end{eqnarray}
the time-dependent physical state of the system is obtained ($\hbar= 1$ is considered) 
\begin{eqnarray}
|\psi(t) \rangle&=&\frac{1}{\sqrt{2N}}\sum_{k}e^{-it\sum_{k}\varepsilon(k) (\beta_{k}^{\dagger} \beta_{k}-\frac{1}{2})}(e^{ikm} \nonumber\\
&+& e^{i(km'+\phi)})a_k^{\dag}|0\rangle~.
\end{eqnarray}

Our interest is calculation of the entanglement between two sites which is measured by the concurrence.  The concurrence between two spins at sites $m$ and $m'$ is obtained by
\begin{equation}
C({\rho _{mm'}}) = \max \{ 0,{\lambda _1} - {\lambda _2} - {\lambda _3} - {\lambda _4}\}, 
\end{equation}
where ${\rho _{mm'}}$ is the reduced density matrix related to any ($m$,$m'$) pair of spins in the chain system,  $\lambda_i$s are square roots of the eigenvalues of the product matrix $R = \sqrt{\rho \widetilde \rho } $ that ${\widetilde \rho _{mm'}} = \left( {\sigma _m^y \otimes \sigma _m'^y} \right)\rho _{mm'}^*\left( {\sigma _m^y \otimes \sigma _m'^y} \right)$ and ${\sigma ^{x,y,z}}$ are  Pauli matrices. The corresponding reduced density matrix $\rho_{m,m'}$, is written as\cite{38,39}
\begin{eqnarray}
\rho_{m,m'}= \left(
             \begin{array}{cccc}
               \langle P_{m}^{\uparrow}P_{m'}^{\uparrow}\rangle & \langle P_{m}^{\uparrow}{\bS}_{m'}^{-}\rangle & \langle {\bS}_{m}^{-}P_{m'}^{\uparrow}\rangle & \langle {\bS}_{m}^{-}{\bS}_{m'}^{-}\rangle \\
               \langle P_{m}^{\uparrow}{\bS}_{m'}^{+}\rangle & \langle P_{m}^{\uparrow}P_{m'}^{\downarrow}\rangle & \langle {\bS}_{m}^{-}{\bS}_{m'}^{+}\rangle & \langle {\bS}_{m}^{-}P_{m'}^{\downarrow}\rangle \\ \langle {\bS}_{m}^{+}P_{m'}^{\uparrow}\rangle & \langle {\bS}_{m}^{+}{\bS}_{m'}^{-}\rangle & \langle P_{m}^{\downarrow}P_{m'}^{\uparrow}\rangle & <P_{m}^{\downarrow}{\bS}_{m'}^{-}\rangle \\
               \langle {\bS}_{m}^{+}{\bS}_{m'}^{+}\rangle & \langle {\bS}_{m}^{+}P_{m'}^{\downarrow}\rangle & \langle P_{m}^{\downarrow}{\bS}_{m'}^{+}\rangle & \langle P_{m}^{\downarrow}P_{m'}^{\downarrow}\rangle\\
               \end{array}\nonumber
               \right),
\label{density matrix1}
\end{eqnarray}
where $P^{\uparrow}= \frac{1}{2}+{\bS}^{z}, P^{\downarrow}= \frac{1}{2}-{\bS}^{z}$ and ${\bS}^{\pm}= {\bS}^{x}\pm i{\bS}^{y}$. Applying Jordan-Wigner transformation, the reduced density matrix will be rewritten as\cite{40}
\begin{eqnarray}
\rho_{m,m'}=\left(
\begin{array}{cccc}
X_{mm'}^{+} & 0 & 0 &f_{m,m'}^{*} \\
0 & Y_{mm'}^{+} & Z_{mm'}^{*} & 0 \\
0 & Z_{mm'} & Y_{mm'}^{-} & 0 \\
f_{m,m'} & 0 & 0 & X_{mm'}^{-} \\
\end{array}
\right). \label{density matrix2}
\end{eqnarray}
While $X^{+}_{mm'}=\langle n_{m} n_{m'}\rangle$, $X^{-}_{mm'}=\langle 1-n_m- n_{m'}+n_m n_{m'}\rangle$, $Y^{+}_{mm'}=\langle n_m(1-n_{m'})\rangle$, 
$Y^{-}_{mm'}=\langle n_{m'}(1-n_{m})\rangle$ and $Z_{mm'}=\langle a_{m}^{\dagger}a_{m'} \rangle$, these phrases can be simplified as
\begin{figure}
\centerline{\psfig{file=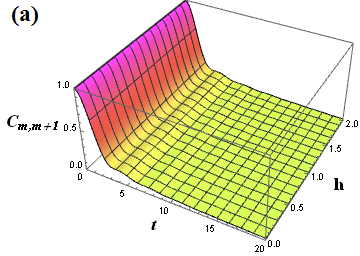,width=1.6 in} \psfig{file=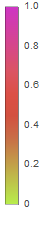,width=0.18 in} \psfig{file=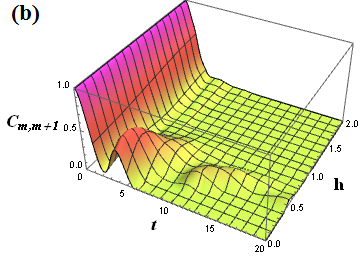,width=1.6 in} \psfig{file=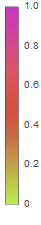,width=0.18 in}}
\centerline{\psfig{file=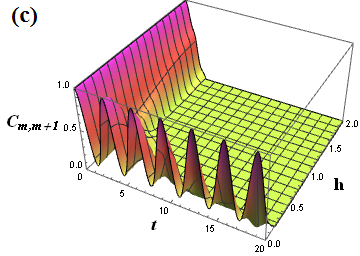,width=1.6 in} \psfig{file=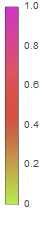,width=0.18 in} \psfig{file=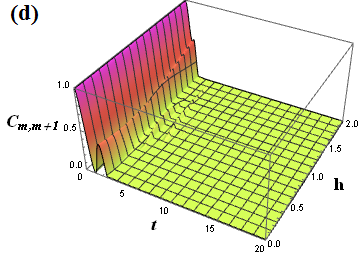,width=1.6 in} \psfig{file=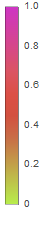,width=0.18 in}}
\centerline{\psfig{file=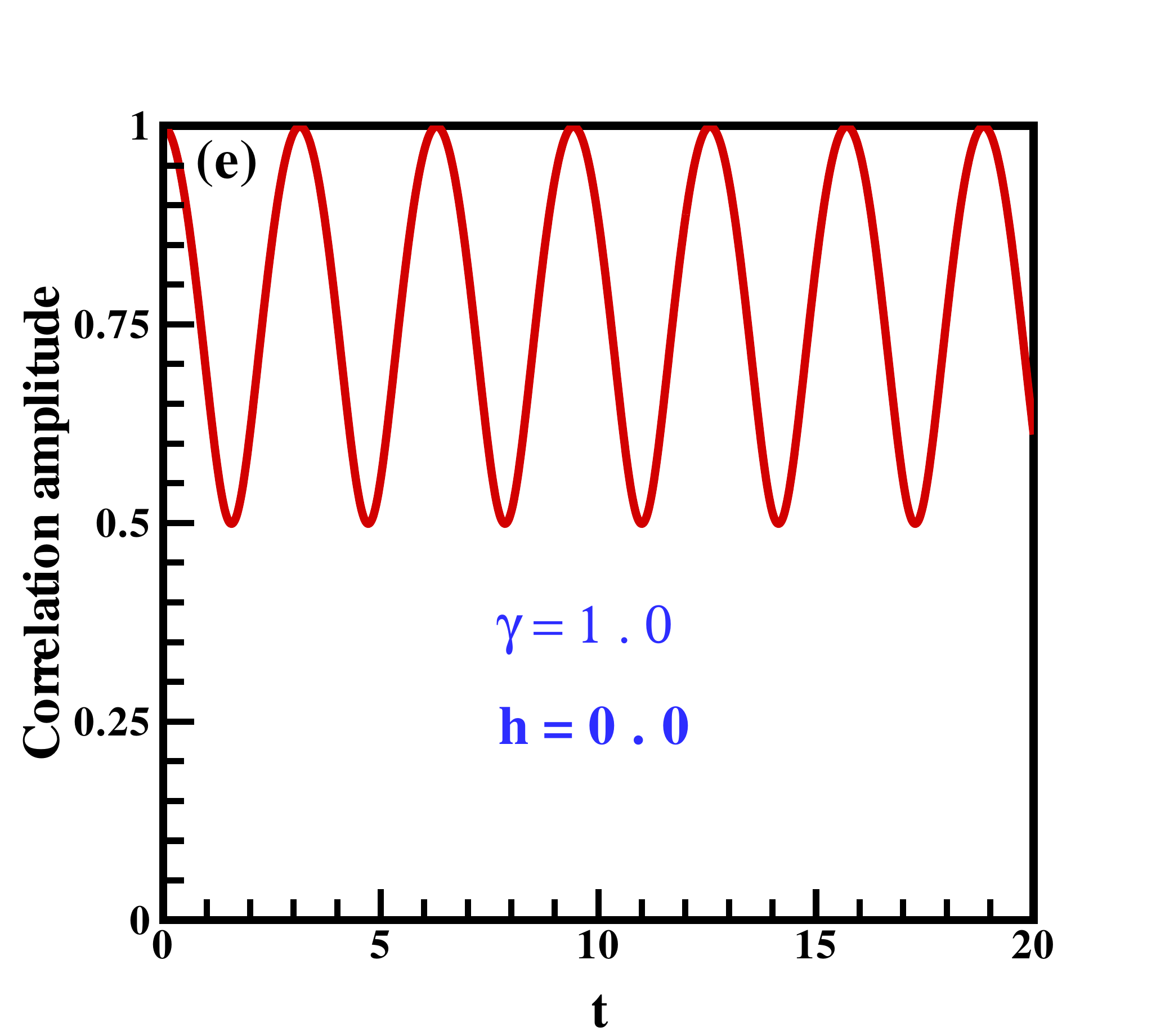,width=1.9 in}}
\caption{Concurrence between pair spins of the system ($C_{m, m+1}$) as a function of time and  magnetic field (a) $\gamma=0$, (b) $\gamma=0.5$, (c) $\gamma=1$, and (d) $\gamma=2$. (e) The correlation amplitude as a function of time for the Ising model in the absence of the transverse field. }
\label{Fig2}
\end{figure}
\begin{eqnarray}
X^{+}_{mm'}&=&\langle a^{\dag}_{m} a_{m}\rangle \langle a^{\dag}_{m'} a_{m'}\rangle+\langle a^{\dag}_{m} a_{m'}\rangle \langle a_{m} a^{\dag}_{m'}\rangle, \nonumber\\
X^{-}_{mm'}&=&X^{+}_{mm'}-\langle a^{\dag}_{m'} a_{m'}\rangle-\langle a^{\dag}_{m} a_{m}\rangle+1, \nonumber\\
Y^{+}_{mm'}&=&X^{+}_{mm'}-\langle a^{\dag}_{m} a_{m}\rangle, \nonumber\\
Y^{-}_{mm'}&=&X^{+}_{mm'}-\langle a^{\dag}_{m'} a_{m'}\rangle \nonumber\\
f_{m,m'} &=&\langle a_{m}^{\dag} (1 - 2 a_{m+1}^{\dag} a_{m+1})\nonumber \\
&\cdots&(1 - 2 a_{m'-1}^{\dag} a_{m'-1}) a_{m'}^{\dag}\rangle.
\end{eqnarray}
In the mentioned relations, ${n_m} = a_m^\dag {a_m}$ is in principle the occupation number operator. Finally, the concurrence between two spins is given by 
\begin{eqnarray}
C_{m,m'}&=&\mbox{max}\{0,\Lambda_{1},\Lambda_{2}\},\nonumber \\
\Lambda_{1}&=&2(|Z_{mm'}|-\sqrt{X^{+}_{mm'}X^{-}_{mm'}}),\nonumber \\
\Lambda_{2}&=&2(|f_{mm'}|-\sqrt{Y^{+}_{mm'}Y^{-}_{mm'}}).
\end{eqnarray}

\section{Results}

In this section, we present our results about the effect of the transverse magnetic field on the behavior of the entanglement. 
To achieve this purpose, the study of different parts of the chain is required: the behavior of the entanglement between the pair of spins in the system ($m$, $m+1$), the edge ($m+1$, $m+2$) and a pair spins of the environment ($m+5$, $m+6$). Fig.~\ref{fig1} shows the symbolic shape of the two-spin system and its environment.  On the other hand, it was stated that specific values of $\gamma$ show different models so that cover all the following models. We consider four values of $\gamma$ as:\\
\\
(a) Symmetric $XY$ model, $\gamma=0$,
\\
(b) Asymmetric ferromagnetic-ferromagnetic (FM-FM) $XY$ model, $0<\gamma<1$,
\\
(c) Ising model, $\gamma=1$, 
\\
(d) Asymmetric ferromagnetic-antiferromagnetic (FM-AFM) $XY$ model, $\gamma>1$.\\ 

In Fig.~\ref{Fig2}, entanglement of the system ($m$, $m+1$) has been studied as a function of time and the magnetic field. In Fig.~\ref{Fig2}~(a), in the symmetric $XY$ model ($\gamma=0$), the amount of entanglement decreases by passing time and applying the magnetic field does not affect on this behavior. By applying $\gamma$ (Fig.~\ref{Fig2}~(b)) and for the asymmetric FM-FM $XY$ model, in the absence of the magnetic field, entanglement decreases instantly and increases after a short time. At the special value of $\gamma=1$, where the model is known as the Ising, the entanglement coming back completely and this behavior occurs alternately (Fig.~\ref{Fig2}~(c)) in the absence of the transverse field.  In Fig.~\ref{Fig2}~(e), we have plotted the overlap between the time-evolved state $|\psi(t) \rangle$ of the pure Ising model ($h=0$, $\gamma=1$) and the initial state $|\psi(0) \rangle$ as a function of time. As is seen, in certain times the time-evolved state will be exactly the same with the initial state. For this reason, the concurrence of the spins in the system appears periodically for the pure Ising model as is clearly seen in Fig.~\ref{Fig2}~(c).   In fact, when the entanglement in the system decreases, it will propagate from the pair of spins in the system to the remaining spins in the chain and when the entanglement in the system increases again, the return of entanglement from the spins in the environment to the system will be observed. We increased the value of $\gamma$, at $\gamma=2$ where the model is known as FM-AFM $XY$, the return of entanglement decreases (Fig.~\ref{Fig2}~(d)). It is noteworthy that, as predicted, with the increase of the magnetic field, for all values of $\gamma$, $C_{m, m+1}$ will be zero. So far, entanglement has been investigated between the pair spins of the system. 

\begin{figure}
\centerline{\psfig{file=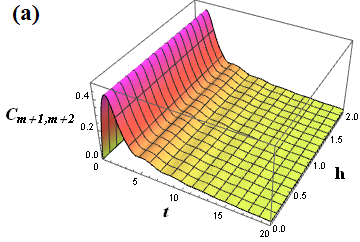,width=1.55 in} \psfig{file=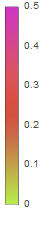,width=0.18 in} \psfig{file=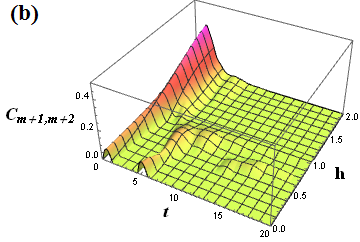,width=1.55 in} \psfig{file=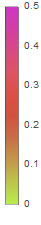,width=0.18 in}}
\centerline{\psfig{file=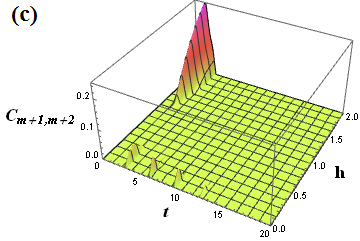,width=1.55 in} \psfig{file=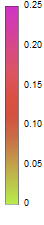,width=0.18 in} \psfig{file=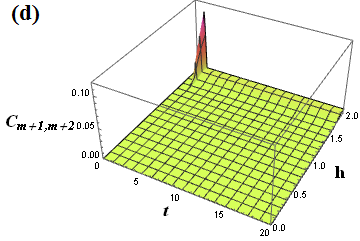,width=1.55 in} \psfig{file=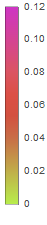,width=0.18 in}}
\caption{Concurrence between pair spins on edge ($C_{m+1, m+2}$) as a function of time and magnetic field (a) $\gamma=0$, (b) $\gamma=0.5$, (c) $\gamma=1$, and (d) $\gamma=2$.}
\label{Fig3}
\end{figure}

Now, we want to know what the behavior of entanglement will be on the edge spins ($m+1$, $m+2$). At the edge, one spin is in the system and another one is in the environment. In fact, the edge is the connection between the system and the environment. It was interesting to us whether there was a difference between the behavior of the entanglement at the edge and the rest of the pair of spins in the chain. In Fig.~\ref{Fig3}, entanglement for edge spins has been perused as a function of  time and the transverse magnetic field .  It is clear from the initial state, there is no entanglement between the pair of spins in the edge at $t=0$. At $\gamma=0$ where the model is the symmetric $XY$,  the entanglement on the edge increases by passing time and shows a peak at short time (Fig.~\ref{Fig3}~(a)). By more increasing time and as $t\longrightarrow \infty$ the entanglement behaves asymptotically independent of the value of the transverse field. In the case of asymmetric FM-FM $XY$ model, a small amount entanglement is generated between spins on the edge for the values of the transverse magnetic fields $h<h_c=1.0$ (where $h_c=1$ is the critical point of the quantum phase transition in this model) is seen in a short time (Fig.~\ref{Fig3}~(b)). But in the region $h>h_c=1.0$, the amount of the short-time induced entanglement increases by increasing the transverse magnetic field. Interesting results are seen in Fig.~\ref{Fig3}~(c) when we considered Ising model, $\gamma=1$. In the region $h<h_c=1.0$, no entanglement is created between spins on the edge. As soon as the transverse magnetic field increases from $h_c=1.0$, a peak is appeared in a short time. Finally,  in the case of asymmetric FM-AFM $XY$ model (Fig.~\ref{Fig3}~(d)), only edge spins will be entangled for enough large transverse magnetic fields after a  very short time. 

\begin{figure}
\centerline{\psfig{file=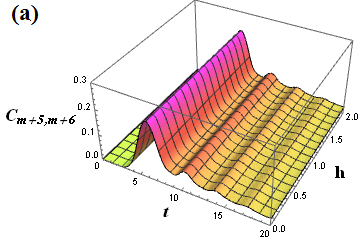,width=1.6 in} \psfig{file=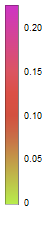,width=0.18 in} \psfig{file=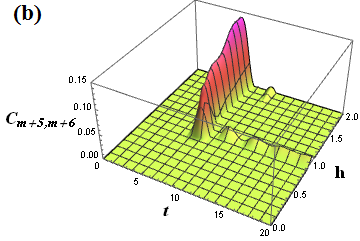,width=1.6 in} \psfig{file=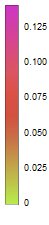,width=0.20 in}}
\centerline{ \psfig{file=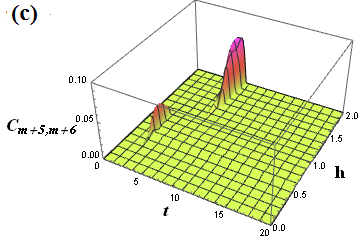,width=1.6 in} \psfig{file=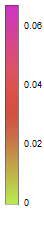,width=0.19 in} \psfig{file=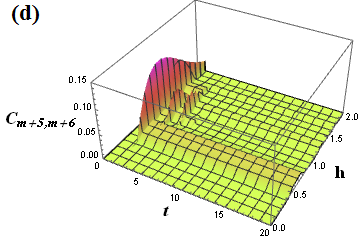,width=1.6 in} \psfig{file=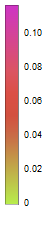,width=0.19 in}}
\caption{Concurrence between a pair spins in the environment ($C_{m+5, m+6}$) as a function of time and magnetic field (a) $\gamma=0$, (b) $\gamma=0.5$, (c) $\gamma=1$, and (d) $\gamma=2$.}
\label{Fig4}
\end{figure}

Heretofore, the behavior of entanglement in the system and the edge was studied and the differences between them were observed. Now, we are going to study the behavior of entanglement between a pair spins of the environment. To do this, a pair of spins were selected as the sample ($m+5$, $m+6$).  In Fig.~\ref{Fig4}, entanglement  in the environment has been studied as a function of time and the transverse magnetic field. At $\gamma=0$ where the model is known as the symmetric XY model, entanglement between the mentioned pair of spins will be created by passing a considerable time (Fig.~\ref{Fig4}~(a)). Transverse magnetic field has no effect in this case.  As time passes, a peak is observed and after a while, the entanglement decreases, but it does not disappear completely. For the asymmetric FM-FM $XY$ model (Fig.~\ref{Fig4}~(b)), only in the region $h>h_c=1.0$ the mentioned pair of spins in the environment will be entangled after a considerable time. At $\gamma=1$, the different behavior is seen for the Ising model (Fig.~\ref{Fig4}~(c)). When the transverse field is close to the critical $h_c=1$, in a short period pair of spins in the environment will be entangled.  In addition in the region $h>h_c$ and for enough large fields, again in a short period the entanglement is seen between mentioned pair spins of the environment. Finally, in the case of asymmetric FM-AFM $XY$ model, in the region $h>h_c=1.0$ the velocity of the propagation of the entanglement into the environment is very high and sooner than other cases the mentioned pair of spins will be entangled (Fig.~\ref{Fig4}~(d)).

\section{CONCLUSION}

In this paper, we have studied the dynamics of entanglement of an anisotropic spin-1/2 XY chain model in the presence of a transverse magnetic field. The behavior of entanglement in the system, edge, and environment has been investigated. To achieve this purpose, we selected a pair of spins in the system (m, m+1), the edge (m+1, m+2), and the environment (m+5, m+6). The physical state of the chain is defined so that the pair of spins in the system are entangled at $t=0$, while the rest of the pair of spins are not entangled. 

In the following, four values of $\gamma$ are selected such that each of them shows one model. Our observations have shown that in the symmetric $XY$ Heisenberg model, the entanglement between the pair of spins in the system decreases exponentially and it can transfer to the rest of the spins. At $\gamma=1$, which shows the Ising model, the entanglement comes back to the system. On the other hand, it is our interest to examine the behavior of entanglement at the edge. The pair of spins on the edge will become entangled immediately and it will be disappeared in a short time. In the next step, a pair of spins was chosen to study the behavior of entanglement in the environment.
We can see that these spins will be entangled after a notable time which is expected since the selected pair of spins is far from the system and the transmission of entanglement from the system to the environment takes time. It's very important to be aware that the entanglement can only be seen under certain conditions on the edge and the environment.

\section{Acknowledgments}

It is our pleasure to thank T. Mohammad Ali Zadeh.\\ 

\section{Appendix} 
Here we want to express one part of our calculations. We calculate  $\langle a^{\dag}_{m}(t) a_{m}(t)\rangle$ as an example.
\begin{eqnarray}
\langle a^{\dag}_{m}(t) a_{m}(t)\rangle = \langle\psi_0|e^{iHt}a^{\dag}_{m}e^{-iHt}e^{iHt}a_{m}e^{-iHt}|\psi_0\rangle= \nonumber\\
\frac{1}{\sqrt{N}}\sum _{k}e^{i(k_1-k_2)m}\langle\psi_0|e^{iHt}a^{\dag}_{k_1}e^{-iHt}e^{iHt}a_{k_2}e^{-iHt}|\psi_0\rangle. \nonumber\\
\label{N_m}
\end{eqnarray}
Using the Bogoliubov operators, Eq. (\ref{N_m}) can be written as follows
\begin{eqnarray}
&\langle& a^{\dag}_{m}(t) a_{m}(t)\rangle =\frac{1}{\sqrt{N}}\sum _{k}e^{i(k_1-k_2)m}\nonumber\\
&\langle&\psi_0|(cos(k_1)e^{it\varepsilon(k_1)}\beta_{k_1}^{\dagger}\nonumber \\
&+&i sin(k_1)e^{-it\varepsilon(-k_1)}\beta_{-k_1}) \nonumber\\
&(&cos(k_2)e^{-it\varepsilon(k_2)}\beta_{k_2}^{\dagger}\nonumber \\
&-&isin(k_2)e^{it\varepsilon(-k_2)}\beta_{-k_2}^{\dagger})|\psi_0\rangle. \nonumber \\
\label{N1_m}
\end{eqnarray}
Finally, the above relation is simplified as 
\begin{eqnarray}
\langle a^{\dag}_{m}(t) a_{m}(t)\rangle &=& \frac{1}{2 N^2}\sum _{k,k'}(1+e^{i(k+\phi)}+e^{-i(k'+\phi))}\nonumber \\
&+&e^{i(k-k')})\nonumber\\
&(&e^{-it(\varepsilon(k)-\varepsilon(k'))}cos^2(k)cos^2(k')\nonumber \\
&+&e^{it(\varepsilon(k)+\varepsilon(k'))}sin^2(k)cos^2(k')\nonumber\\
&+&e^{-it(\varepsilon(k)+\varepsilon(k'))}sin^2(k')cos^2(k)\nonumber\\
&+& e^{it(\varepsilon(k)-\varepsilon(k'))}sin^2(k)sin^2(k')\nonumber\\
&+&\frac{1}{4}sin(2k)sin(2k')(-e^{it(\varepsilon(k)-\varepsilon(k'))}\nonumber\\
&+&e^{it(\varepsilon(k)+\varepsilon(k'))}\nonumber\\
&+&e^{-it(\varepsilon(k)+\varepsilon(k'))}-e^{-it(\varepsilon(k)-\varepsilon(k'))}))\nonumber\\
&+& \frac{1}{4 N^2}\sum _{k,k'}(sin^2(2k') (1+cos(k+\phi))\nonumber\\
&(&2-e^{2it(\varepsilon(k')}-e^{-2it(\varepsilon(k')}). 
\end{eqnarray}
\vspace{0.3cm}


\end{document}